\documentclass[%
 aip,
 amsmath,amssymb,
 reprint,%
cha]{revtex4-1}

\usepackage{graphicx}
\usepackage{dcolumn}
\usepackage{bm}

\usepackage[utf8]{inputenc}
\usepackage[T1]{fontenc}
\usepackage{mathptmx}

\usepackage{color}

\begin{document}

\preprint{AIP/123-QED}

\title[The Winfree model with non-infinitesimal PRC]{The Winfree model with non-infinitesimal phase-response curve: Ott-Antonsen theory}

\author{Diego Paz\'o}
 \email[Author to whom correspondence should be addressed: ]{pazo@ifca.unican.es}
 \affiliation{Instituto de F\'isica de Cantabria (IFCA), CSIC-Universidad de
 Cantabria, 39005 Santander, Spain}
\author{Rafael Gallego}%
\affiliation{Departamento de Matem\'aticas, Universidad de Oviedo, Campus de
Viesques, 33203 Gij\'on, Spain}%

\date{\today}

\begin{abstract}
A novel generalization of the Winfree model of globally coupled phase oscillators,
representing phase reduction under finite coupling, is studied analytically.
We consider interactions through a non-infinitesimal (or finite) phase-response
curve (PRC), in contrast to the infinitesimal PRC of the original model.
For a family of non-infinitesimal PRCs,  the global dynamics is captured
by one complex-valued ordinary differential equation resorting to the Ott-Antonsen ansatz.
The phase diagrams are thereupon obtained for four illustrative cases of non-infinitesimal PRC.
Bistability between collective synchronization and full desynchronization is observed in all cases.
\end{abstract}

\maketitle

\begin{quotation}
In 1967 Winfree proposed a model for the spontaneous synchronization of
large ensembles of biological oscillators \cite{Win67}.
The Winfree model played a seminal role in the field 
of collective synchrony, inspiring the Kuramoto
model \cite{Kur75,Str00} as well as promoting recent advances
in theoretical neuroscience \cite{MP18}.
In spite of the simplifying assumptions of the Winfree model,
uniform all-to-all weak coupling, analytical solutions have been 
found only recently using the Ott-Antonsen ansatz\cite{PM14,gallego17}, see also \cite{AS01}.
Weak coupling is implicit in the use of phase oscillators as the units of the model.
Moreover, their interactions are modeled by the so-called infinitesimal phase-response curve (iPRC),
which is only valid  in the limit of vanishing coupling. In this paper we extend the Winfree model 
considering a non-infinitesimal (also called finite) PRC, such that the phase shift 
of one oscillator is not proportional to the magnitude of the input.
For a family of non-infinitesimal PRCs, and a Lorentzian distribution 
of natural frequencies, the global dynamics is captured by one complex-valued ordinary differential equation
by means of the Ott-Antonsen ansatz \cite{OA08,OA09,OHA11}.
We obtain the phase diagrams for four instructive cases.
\end{quotation}

\section{Introduction}
Collective synchronization in large ensembles of self-sustained oscillators is a pervasive 
phenomenon in nature and technology \cite{Win80,PRK01,Izh07}.
The first successful attempt to model collective synchronization is due to Winfree \cite{Win67}.
Relying on his intuition he devised a model where the only degrees of freedom
were the oscillators' phases, and the coupling was uniform and global (i.e.~mean-field type).
In the numerical simulations a macroscopic cluster of synchronized oscillators 
emerged spontaneously when either the natural frequencies of the oscillators were {\color{black} narrowly} distributed or the coupling
was large enough. In mathematical language, the phases in the Winfree model 
are governed by a set of $N$ ordinary differential equations ($i=1,\ldots,N$):
\begin{subequations} \label{winfree} 
\begin{eqnarray}
\dot\theta_i=\omega_i+ {\tilde Q}(\theta_i)\, A, \label{wina}\\
A= \frac{\epsilon}{N} \sum_{j=1}^N P(\theta_j). \label{winb}
\end{eqnarray}
\end{subequations}
Here $\omega_i$ is the natural frequency of the $i$-th oscillator, and
$\epsilon>0$ is a parameter controlling the coupling strength. The $2\pi$-periodic function $P$ specifies the
pulse shape. The function $\tilde Q$ is also $2\pi$-periodic and 
is either called infinitesimal (or linear) phase-response curve (iPRC), or
sensitivity function \cite{Win80,Kur84,Izh07}.

As already mentioned, the Winfree model relies on two assumptions: weak coupling
and all-to-all geometry.  Weak coupling permits, first of all, ignoring the
oscillators' amplitudes: the limit cycles are strongly attracting 
compared to perturbations causing amplitudes to be strongly damped degrees of freedom.
In addition, the effect of the mean field
$A$  on the phase is exactly proportional to $A$ ---higher powers of $A$ are absent in
Eq.~\eqref{wina}---, {\color{black} which} only holds in the limit of asymptotically small
interactions \cite{Win80,Kur84,Izh07,sacre14,pietras19}.

In this work we generalized the Winfree model considering nonlinear interactions. Mathematical
tractability imposes certain restrictions on the distribution of the natural frequencies and
on the class of ``non-infinitesimal'' (also called ``finite'' or ``non-linear'') PRCs, 
but we believe it is remarkable that such analytic solutions exist. 
This limited progress should be welcome given the relevance of PRC theory 
in theoretical neuroscience \cite{ET10,Boergers},
and recent experiments evidencing the {\color{black} insufficiency} of 
the linear approximation \cite{rode19,totz}.
Our analysis is based on the so-called ``Ott-Antonsen (OA) theory'', which assumes 
a certain ansatz (the Poisson kernel) for the density of the phases
in the thermodynamic limit ($N\to\infty$). 
The OA ansatz was initially applied to the Kuramoto model and its variants \cite{OA08,OA09},
but eventually found application in several systems of pulse-coupled oscillators: the original Winfree model 
\cite{PM14,gallego17} (and a variant with heterogeneous iPRCs \cite{PMG19}),
ensembles of theta neurons \cite{LBS13,laing14,SLB14},
quadratic integrate-and-fire neurons \cite{MPR15,PM16,RP16}, and excitable active rotators \cite{okeeffe16,roulet16}. 

\section{Winfree model with non-infinitesimal PRC}

We consider a modification of the Winfree model \eqref{winfree}, in which Eq.~\eqref{wina}
is replaced by 
\begin{equation}
\dot\theta_i=\omega_i+ Q(\theta_i,A),\qquad i=1,\ldots,N,
\label{model} 
\end{equation}
where $A$ is the mean field defined by \eqref{winb}.
At the lowest order in $A$, the model \eqref{model}
converges to the Winfree model \eqref{winfree}:
$dQ(\theta_i,A)/dA|_{A=0}=\tilde Q(\theta_i)$.
{\color{black} Assuming $Q(\theta,A)$ linear in $A$ is equivalent to approximate the isochrons
of a limit cycle by straight lines (or hyperplanes if the dimensionality is larger than two) 
in the phase reduction procedure \cite{Kur84,Izh07}.
}

\subsection{Non-infinitesimal PRC}

Prior to specifying the PRC $Q$, we devote a few lines to the
iPRCs. Traditionally, iPRCs are classified as type I or type II \cite{hansel95}. For type II,  either
an advance or a delay in the phase are possible depending upon the timing of the
perturbation, while in the case of type I the timing of the perturbation does
not change the sign of the phase shift.  The canonical examples of each
type\cite{Izh07,sacre14} are $\tilde Q(\theta)\propto1-\cos\theta$ for type I (e.g. the theta neuron), and
$\tilde Q(\theta)\propto\sin\theta$ for type II (e.g. the Stuart-Landau oscillator).
For non-infinitesimal PRCs, the previous classification falls short as the
character of $Q$ may change with the strength of the stimulus \cite{Izh07,sacre14}. 

The types of PRC we consider are conditioned by the applicability of the OA ansatz,
as it enables a drastic dimensionality reduction.
The OA ansatz imposes that no harmonics in $\theta$ beyond the first one 
are present in $Q(\theta,A)$.  Still, the family of PRCs with
only first harmonic in $\theta$ is wide enough to make the problem nontrivial.
As we shall adopt pulses $P(\theta)$ with peak value at $\theta=0$ (and multiples of $2\pi$), 
we impose the additional constraint $Q(0,A)=0$ motivated by the fact that the PRC 
vanishes at spiking/flashing times for most neurons \cite{reyes93,netoff05} 
and certain fireflies \cite{buck88,hanson78}.
Therefore, we restrict to a family of PRCs of this form:
\begin{equation}
 Q(\theta,A)=f_1(A) (1-\cos\theta) - f_2(A) \sin\theta,
\label{PRC}
 \end{equation}
where $f_1$ and $f_2$ are arbitrary functions of $A$,
provided that $f_{1,2}(0)=0$ for obvious physical reasons.
In similarity with the classification of iPRCs, we refer to the two terms in \eqref{PRC}, proportional to $(1-\cos\theta)$
and $\sin\theta$, as the type-I and the type-II components of the PRC, respectively.

\subsection{Pulse shape}

\begin{figure}
\centerline{\includegraphics[width=75mm,clip=true]{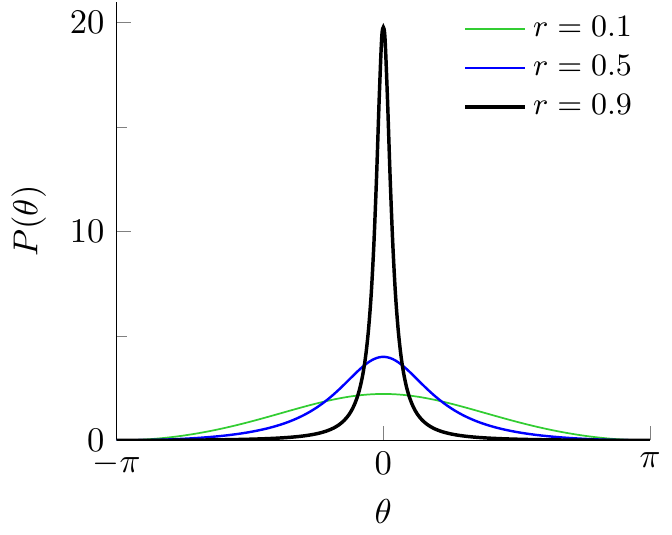}}
\caption{Rectified-Poisson pulse \eqref{eq:pulse} for three values of parameter
$r$.}
\label{fig:pulse}
\end{figure}

In the study of the classical Winfree model several pulse shapes can be considered, see \cite{gallego17}.
In this work, we adopt a ``rectified Poisson kernel''\cite{gallego17}:
\begin{equation}
P(\theta)=\frac{(1-r)(1+\cos\theta)}{1-2r\cos\theta +r^2} .
\label{eq:pulse}
\end{equation}
This is a particularly convenient shape for the theoretical analysis below.  $P(\theta)$ is a
symmetric unimodal function in the interval $[-\pi,\pi]$ (with the normalization
$\int_{-\pi}^{\pi} P(\theta) d\theta=2\pi$) that peaks at $\theta=0$ and
vanishes at $\theta=\pm\pi$.  Parameter $r$ is a real number allowing a
continuous interpolation between a flat pulse for $r=-1$ and a Dirac-delta
pulse, $P(\theta)=2\pi\delta(\theta)$, for $r=1$. In Fig.~\ref{fig:pulse} the
pulse function $P(\theta)$ is depicted for three different values of $r$.

\subsection{Natural frequencies}

For the sake of achieving the maximal dimensionality reduction, we assume
the natural frequencies  to be distributed according to a Lorentzian distribution  
of half-width $\Delta$ centered at $\omega_0$:
\begin{equation}
g(\omega)= \frac{\Delta/\pi}{(\omega-\omega_0)^2+\Delta^2}.
\label{lorentzian}
\end{equation}

\section{Ott-Antonsen theory}
Once the building blocks of the model have been introduced,
we apply the OA theory~\cite{OA08}. In this way 
we derive a complex-valued ODE reproducing the long-time evolution of the model 
at the macroscopic level. As the procedure is standard\cite{OA08,gallego17}, the readers interested in the
final result are pointed to Eqs.~\eqref{eq:Z} and \eqref{hZ}.
 
First of all, one must realize that our model \eqref{model} belongs to a general class 
of oscillator systems of the form
\begin{gather}\label{eq:winsys}
  \dot\theta_i(t)=\omega_i+B(t)+\mathrm{Im}\left[H(t)e^{-i\theta_i(t)}\right] ,
\end{gather}
which can be analyzed with the OA ansatz \cite{OA08,OA09,OHA11,PD16}.
Functions $B$ and $H$ may depend explicitly on time or indirectly through a mean field.
For the PRC \eqref{PRC} we have 
\begin{equation}\label{BH}
B(t)=f_1(A), \qquad H(t)=f_2(A)-if_1(A). 
\end{equation}

In the thermodynamic limit we can define a phase density $F(\theta|\omega,t)$,
such that $F(\theta|\omega,t)d\theta$ is the fraction of oscillators of frequency $\omega$
at time $t$, with phases in the interval $[\theta,\theta+d\theta]$.
It is convenient to introduce the Fourier expansion of the density
\begin{gather}
  F(\theta|\omega,t)=\sum_{m=-\infty}^\infty\alpha_m(\omega,t)e^{im\theta} ,\nonumber
\end{gather}
with $\alpha_{-m}=\alpha_m^*$.
We notice as well that, by conservation of the number {\color{black} of} oscillators, $F$ satisfies the continuity 
equation:  $\partial_tF+\partial_\theta(F\dot\theta)=0$,
where $\dot\theta$ is the speed of an oscillator of natural frequency $\omega$.
Inserting the Fourier series of $F$ into the continuity equation we get:
\begin{equation}\label{am}
  \partial_t\alpha_m(\omega,t)
  =-im(\omega+B)\alpha_m+
  \frac{m}{2}\left(H^*\alpha_{m-1}-H\alpha_{m+1}\right). 
\end{equation}
A particular solution of this equation, the OA ansatz, is obtained 
equating the coefficient of $m$-th mode to the $m$-th power of the first mode:
$\alpha_m=\alpha_1^m$.
Hence, for the solution in this so-called OA manifold, 
we only need to consider the evolution of $\alpha_1\equiv\alpha$:
\begin{equation}\label{eq:alpha}
  \partial_t\alpha(\omega,t)
  =-i(\omega+B)\alpha+
  \frac{1}{2}\left(H^*-H\alpha^2\right). 
\end{equation}
This is still an infinite set of coupled integro-differential equations.
A sharp reduction in the dimensionality of the problem is achieved for rational $g(\omega)$,
and specially for the Lorentzian distribution \cite{OA08}. 
As the Kuramoto order parameter\cite{Kur75} $Z=\overline{e^{i\theta}}$ is related to $\alpha$ via $Z^*(t)=\int_{-\infty}^\infty
\alpha(\omega,t) g(\omega) d\omega$, we can evaluate this integral resorting to the {\color{black} residue} theorem obtaining
$Z^*(t)=\alpha(\omega_0-i\Delta,t)$. (This is the result of performing an analytic continuation of $\alpha$ from real to
complex $\omega$, and evaluating $\alpha$ at the pole of $g(\omega)$ in the lower half $\omega$-plane.)
Thus, setting $\omega=\omega_0-i\Delta$ in \eqref{eq:alpha}, we get a complex-valued ODE for
the Kuramoto order parameter:
\begin{equation}\label{eq:Z}
  \dot Z=(-\Delta+i\omega_0)Z-\frac{i}2f_1(A)(1-Z)^2+\frac12f_2(A)(1-Z^2),
\end{equation}
where $B$ and $H$ have been written in terms of $f_1$ and $f_2$ according to Eq.~\eqref{BH}.
To close Eq.~\eqref{eq:Z} we need to express the mean field $A$ as a function of $Z$.
For the pulse shape in Eq.~\eqref{eq:pulse} 
and a Lorentzian frequency distribution
it can be proven (see \cite{gallego17} or the supplemental material of \cite{MP18}) that:
\begin{equation}\label{hZ}
A=\epsilon \, \mathrm{Re}\left(\frac{1+Z}{1-r \, Z}\right).
\end{equation}
{\color{black} Note that $0\le A \le A_{\mathrm{max}}$, where the maximal value 
$A_{\mathrm{max}}=2\epsilon/(1-r)$ is achieved if $Z=1$ (all oscillators exactly at $\theta_j=0$).}
In addition to this, the central natural frequency $\omega_0$ is hereafter set to 1, as this can
always be achieved by rescaling time and $f_{1,2}$.

\section{Four illustrative PRC\lowercase{s}}

Among the infinite set of functions $f_1(A)$ and $f_2(A)$ we selected a few illustrative case studies. 
In each of these cases the character of the PRC undergoes a crossover
as $A$ grows: from one iPRC type to a different PRC type for large $A$. 
We denote the limiting PRC at $A\to\infty$ as `asymptotic PRC' (aPRC).
From now on, we apply the classical distinction between types I and II to both 
iPRCs and aPRCs. Recall that if the sign is the same for all $\theta$ we call the iPRC (or the aPRC) type I (implying $f_2=0$),
while in the complementary case with $f_1=0$ we refer to the iPRC (or to the aPRC) as canonical type II,
or simply type II. Notably, type II may either promote or impede synchronization
depending on the sign of $f_2$. In turn, we distinguish between two subclasses
of the type II: $\mathrm{II}_s$ ($f_2>0$)
and $\mathrm{II}_r$ ($f_2<0$) corresponding to the synchronizing and repulsive interactions, respectively.

\begin{table} 
\caption{\label{tabla} The four cases of non-infinitesimal PRCs
analyzed in this paper. Functions $f_1$ and $f_2$ determine the PRC in \eqref{PRC} ($\sigma(A)$ is
a crossover function, see \eqref{sigma}).
The code in the last column indicates the iPRC and the asymptotic PRC at large $A$ (see text).}
\begin{ruledtabular}
\begin{tabular}{cccc}
Case & $f_1(A)$&  $f_2(A)$ & Code\\
\hline
a & $\sigma(A)$ & $A \, \sigma(A)$ & $\mathrm{I}-\mathrm{II}_s$\\
b & $A\,  \sigma(A)$ & $\sigma(A)$& $\mathrm{II}_s-\mathrm{I}$\\
c & $0$ & $(1-A)\sigma(A)$ & $\mathrm{II}_s-\mathrm{II}_r$\\
d & $0$ & $-(1-A)\sigma(A)$ &  $\mathrm{II}_r-\mathrm{II}_s$\\
\end{tabular}
\end{ruledtabular}
\end{table}

\begin{figure}
\centerline{\includegraphics[width=0.99\linewidth,clip=true]{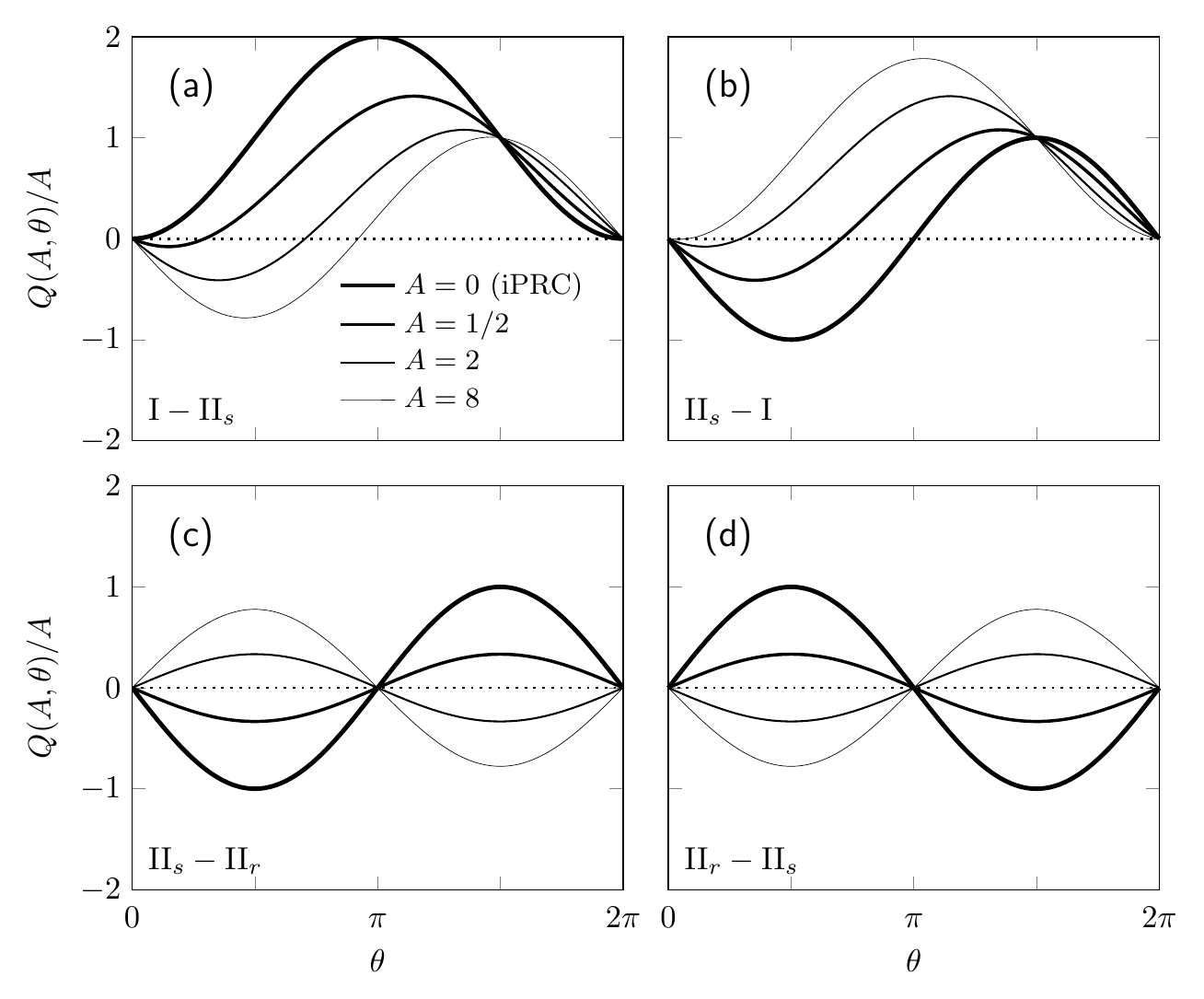}}
\caption{The non-infinitesimal PRCs analyzed in this work as a function of $\theta$
for four representative values of $A$, including $A=0^+$ (iPRC) and $A=8$ (resembling the aPRC). Panels
(a) to (d) correspond to cases a to d, respectively (see Table \ref{tabla}).
The code iPRC-aPRC is indicated in each panel.}
\label{fig:prcs}
\end{figure}

As we are interested in introducing one crossover in the PRC between the iPRC and the aPRC, 
and have three fundamentally different types (I, $\mathrm{II}_s$, and $\mathrm{II}_r$) this gives
6 possible combinations.
However, we shall consider only four of these iPRC-aPRC pairs, since only type $\mathrm{II}_s$ favors synchrony 
and is to be included either in the iPRC or in the aPRC. Otherwise no synchronization phenomena are expected:
type I is neutral and type $\mathrm{II}_r$ is repulsive.
Hence, we  focus on the four cases listed in Table \ref{tabla}, in which different PRC types
characterize small and large $A$ regimes. As a guide, in the 
fourth column of the Table we write a code X-Y, where X refers to the iPRC and Y to the aPRC.
The saturation function $\sigma(A)$ in the Table has positive slope at $A=0$, and saturates at large $A$. 
In particular, we chose this specific saturation function in our study:
\begin{equation}\label{sigma}
 \sigma(A)=\frac{A}{1+A}.
\end{equation}
(Our results have been occasionally tested against another choice $\sigma(A)=\tanh(A)$, finding no qualitative difference.)
Graphical representations of the four PRCs (cases a to d){\color{black}, for four representative $A$ values,}
are shown in Figs.~\ref{fig:prcs}(a)-(d). {\color{black} In each panel the PRC appears divided by $A$ as usual \cite{Izh07},
and the lack of overlapping between different lines evidences its nonlinearity.}

\begin{figure}
\centerline{\includegraphics[width=0.99\linewidth,clip=true]{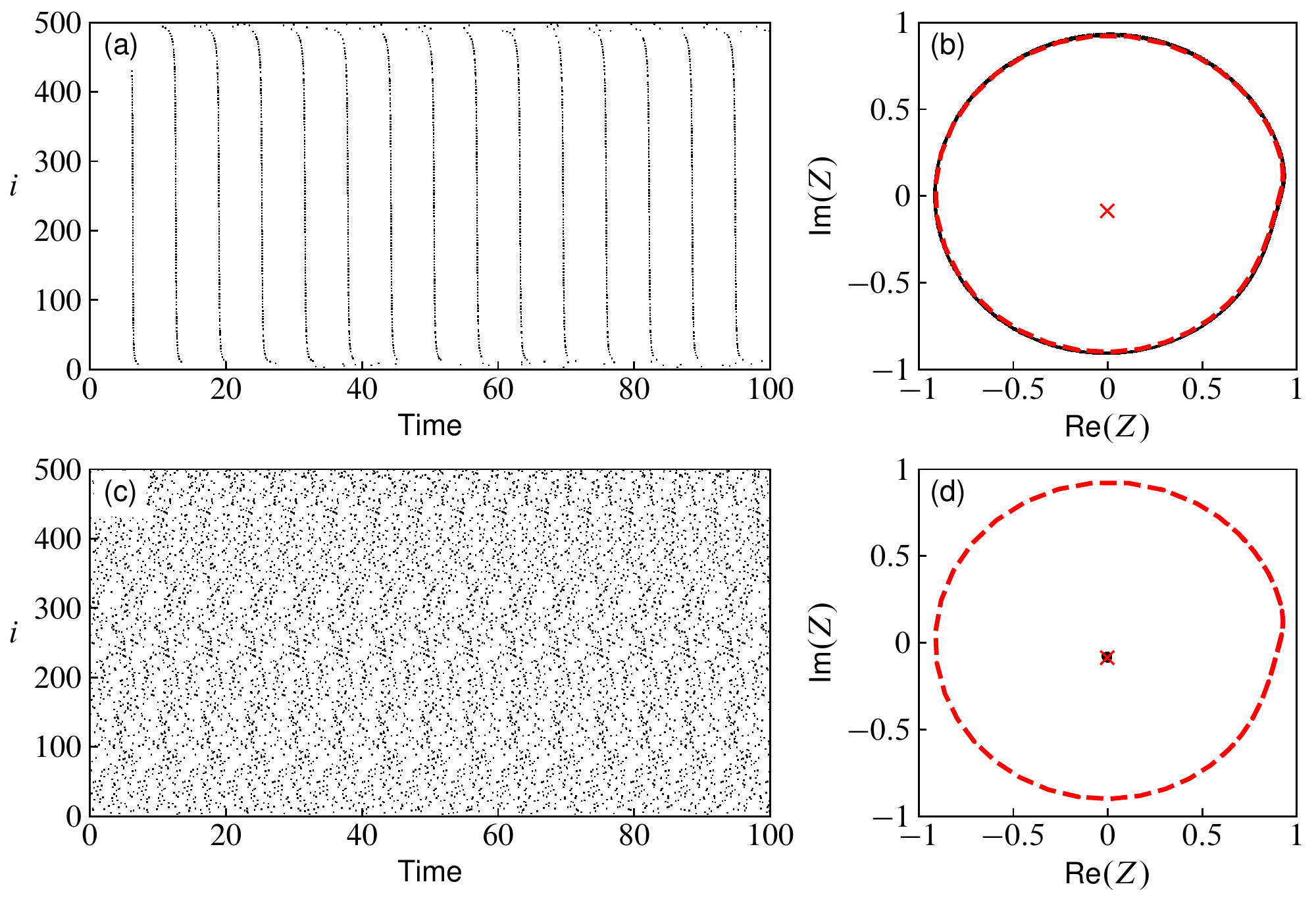}}
\caption{(a) Raster plot ---a dot indicates at which time one oscillator phase crosses a multiple of $2\pi$---
for a population of $N=500$ and the non-infinitesimal PRC of case d, see Table I and Fig.~\ref{fig:prcs}(d).
The initial condition is uniform $\theta_j(t=0)=0.01$ and parameters are $\Delta=0.01$, $r=0.9$,
$\epsilon=0.4$. The frequencies are deterministically drawn from a Lorentzian distribution: 
$\omega_i=\omega_0+\Delta \tan[\pi(2i-N-1)/(2N)]$.
(c) The same as (a) but for a random initial distribution of phases. 
(b) and (d) depict the  Kuramoto order parameter $Z(t)=N^{-1}\sum_j e^{i\theta_j(t)}$ 
for 50 t.u., once the simulations in (a) and (c), respectively,  reached  the stationary state.
The red dashed line and the red cross in panels (b) and (d) are
the periodic and fixed point attractors of Eq.~\eqref{eq:Z}, coexisting at the same parameter values.}
\label{fig:raster}
\end{figure}

In the next section we obtain the phase diagrams corresponding to each of the four cases introduced here,
based on the analysis of the complex-valued ordinary differential equation \eqref{eq:Z}.
But before doing so, it is worth making direct simulations of the full system \eqref{model},
and test (and understand) the correspondence with the solutions of Eq.~\eqref{eq:Z}.
We simulated the full model in case d with $\omega_0=1$, heterogeneity parameter $\Delta=0.01$, 
pulse-shape parameter $r=0.9$ and coupling constant $\epsilon=0.4$.
As may be seen in Fig.~\ref{fig:raster}, the population exhibits bistability between a desynchronized state
and a synchronized state with some oscillators oscillating with the same frequency. This bistability 
is not surprising as the system is ``more synchronizing'' when already synchronized since the aPRC is of type $\mathrm{II}_s$,
while it is hardly synchronizable when already desynchronized by virtue of the type $\mathrm{II}_r$ iPRC.
In terms of $Z$, the synchronous
solution is (approximately) a periodic orbit, while
the desynchronized state
exhibits only small fluctuations around a point due to finite size effects ($N=500$). 
The agreement with the stable fixed point and the stable limit cycle of Eq.~\eqref{eq:Z},
also represented in Figs.~\ref{fig:raster}(b) and \ref{fig:raster}(d), is excellent.

\section{Phase diagrams}

In the {\color{black} remainder} of this paper we obtain the phase diagrams
for the four reference cases by means of Eq.~\eqref{eq:Z}.  As \eqref{eq:Z} is a (generic) planar
system, the only possible attractors are fixed points and limit cycles.
{\color{black} 
Their bifurcation loci, depicted in the phase diagrams below, have been obtained 
using the {\sc matcont} toolbox \cite{matcont} of {\sc matlab}.}
Moreover, we recall that $Z$ is only physically meaningful inside the unit disk
$|Z|\le1$, and therefore attractors and bifurcations occurring outside it are
ignored.  As seen in Fig.~\ref{fig:raster}, limit cycles correspond to synchronized solutions, 
in which a macroscopic part of the population rotates at the same average frequency.

\begin{figure*}
\centerline{\includegraphics[width=0.9\textwidth,clip=true]{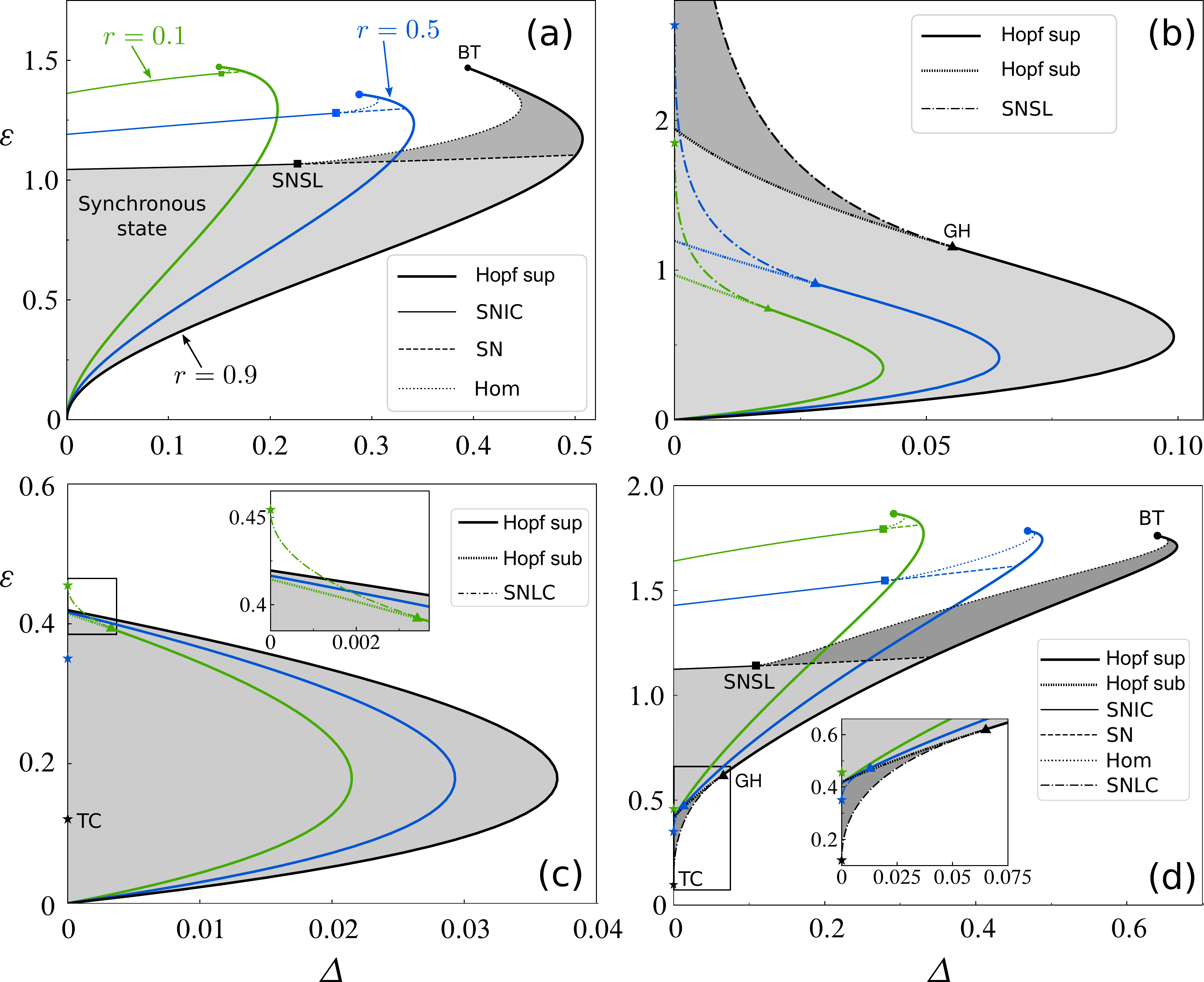}}
\caption{Synchronization regions of the Winfree model in the $(\Delta,\varepsilon)$-plane
for the four cases of PRCs described in Table~\ref{tabla}, and three different
  values of the parameter $r\in\{0.1,0.5,0.9\}$. {\color{black} Panels (a), (b), (c) and (d)
  correspond to cases a, b, c, and d, respectively.}
  For the value $r=0.9$,
  light shaded regions indicate where there is a stable limit cycle, corresponding to a
  macroscopic synchronized state. In the dark shaded regions, the limit cycle
  (synchronous state) coexists with a stable fixed point (asynchronous state).
  The codimension-two points are depicted by specific symbols:
  Generalized Hopf (GH-{$\blacktriangle$}),  saddle-node
  separatrix-loop (SNSL-{$\blacksquare$}), Bogdanov-Takens
  (BT-$\bullet$), and transcritical  (TC-{$\bigstar$}) bifurcations.  
  The bifurcation corresponding to each line type is indicated in the legend of the respective panel.
  {\color{black} Insets in panels (c) and (d) are magnifications of the regions inside the respective rectangles.}}
\label{fig:pd}
\end{figure*}

\subsection{Case a: $\mathrm{I}-\mathrm{II}_s$}

In Fig.~\ref{fig:pd}(a) we show the phase diagram spanned by parameters $\Delta$ and $\varepsilon$.
Bifurcation lines for three values of parameter $r$, controlling the pulse width, are depicted.
The results almost replicate those in Ref.~\cite{gallego17} for the standard Winfree model 
with type $\mathrm{II}_s$ iPRC. Synchronization is
found in two adjacent regions,
in one of them (dark shaded) coexisting with a desynchronized state. 
(There exists a region (not shown) besides the bistability region where two desynchronized
states coexist, see \cite{PM14,gallego17}).
In contrast to the averaging approximation (the Kuramoto-Sakaguchi model), valid at small $\epsilon$ and $\Delta$,
synchronization becomes impossible if the population is too heterogeneous (large $\Delta$).

For small coupling (and heterogeneity), synchronization
emerges from a supercritical Hopf bifurcation 
undergone by the desynchronized state, akin to the classical Kuramoto transition\cite{OA08}. This Hopf bifurcation
line terminates at a double zero eigenvalue (Bogdanov-Takens, BT) point. A homoclinic (Hom)
line emanates from the BT point limiting the coexistence region. 
As observed for the regular Winfree model \cite{PM14,gallego17} 
synchronization is more efficient for narrow pulses.
The pulse width does not qualitatively  change the phase diagram.

The phase diagram only differs appreciably from those in ~\cite{gallego17} at the origin. 
We see that,
due to the type-I iPRC, the Hopf line approaches the origin with an infinite slope.
In particular, the asymptotic dependence of the critical $\epsilon_H$ on $\Delta$ 
{\color{black} follows an unusual square-root law with the frequency dispersion $\Delta$:}
\begin{gather}\label{sqrt}
  \varepsilon_H=\sqrt{\frac{2\Delta}{1+r}}.
\end{gather}
We can deduce this result deriving the associated Kuramoto-Sakaguchi model  
of model \eqref{model} via averaging. Or, alternatively,
preserving in \eqref{eq:Z} only linear, rotationally invariant terms in $Z$, and equating the linear
coefficient to $i\Omega$.

\subsection{Case b: $\mathrm{II}_s-\mathrm{I}$}
In case b, iPRC and aPRC are interchanged with respect to case a. This means that synchronization
is favored at small coupling, but becomes increasingly difficult as the coupling grows.
{\color{black} Accordingly, the 
phase diagram in Fig.~\ref{fig:pd}(b) shows the
expected supercritical Hopf bifurcation line emanating as a straight line
from the origin\cite{PM14,gallego17}: $\epsilon_H\propto\Delta+O(\Delta^2)$. 

At large $\epsilon$ there is
a bistability region 
such that} the synchronized state  
disappears in a saddle-node bifurcation of limit cycles (SNLC).
The locus of the SNLC is a line that emanates from a generalized Hopf (or Bautin) point (GH),
and terminates at the $\epsilon$ axis at a point marked with a star on the $\epsilon$-axis of the phase diagram.
The stars pinpoint the  (equivariant) transcritical (TC) bifurcation \cite{ashwin92}, in which
the fully synchronized state ($\theta_i(t)=\theta_j(t)$) of identical oscillators ($\Delta=0$) becomes unstable.
For $r=0.9$, the instability of full synchronization takes place at $\epsilon_c=9.555\ldots$, 
far above the range  of $\epsilon$ displayed in the phase diagram. 
The location of $\epsilon_c$ was not calculated using \eqref{eq:Z}, but {\color{black} by}
directly looking for the stability threshold to the fully synchronized state, see Appendix.

Finally, note that the synchronization region shrinks as the pulse
becomes wider, but there is not a qualitative change in the phase diagram whatsoever.

\subsection{Case c: $\mathrm{II}_s-\mathrm{II}_r$}

In this case the aPRC is repulsive, in contrast to case b where the aPRC is type I (i.e.~neutral in terms of synchronization).
In turn the phase diagram in Fig.~\ref{fig:pd}(c) shows a quite small synchronization region
(notice the scale of the axes). Synchronization is bounded exclusively by a supercritical Hopf bifurcation,
save for broad pulses. In the latter case a GH point is found, and  
the Hopf bifurcation is subcritical at the left of it. 
{\color{black} Accordingly,}
we find a bistability region bounded by a line of
saddle-node bifurcation of limit cycles (SNLC) and a subcritical Hopf bifurcation{\color{black}, as in case b}.  
The precise value of $r$ below
which the bistability region exists (i.e.~the GH point is present) is
$r_*\simeq0.27891$.

Note also the presence of a TC point in the phase diagram at $\Delta=0$, above which full synchrony destabilizes
\footnote{As the ``OA manifold'' is not attracting for identical oscillators \cite{OA08,OA09,OHA11}, 
the resulting dynamics depends on the initial conditions. (For initial conditions in the OA manifold,
such as purely random phases ($Z=0$), the system converges to a state of
quasiperiodic partial synchronization \cite{Vre96,PR15},
a state which $Z$ oscillates periodically while the individual oscillators behave quasiperiodically.)}.
The transcritical bifurcation is not structurally stable, see e.g. Fig.~11 in \cite{crawford91}, and increasing $\Delta$ from $0$ 
may either leave no trace of bifurcation or ``decay'' into two saddle-node bifurcation of limit cycles.
The latter scenario occurs for $r<0.27577\ldots$, see the bifurcation lines for $r=0.1$ in Fig.~\ref{fig:pd}(c),
but in our case one of the bifurcations is not shown as it entails $|Z|>1$.

\subsection{Case d: $\mathrm{II}_r-\mathrm{II}_s$}

Case d {\color{black} exhibits the most complex phase diagram among all those obtained here.}
The aPRC is of type $\mathrm{II}_s$,
as in case a, and (accordingly) the large $\epsilon$ region 
{\color{black} is organized by two codimension-two points: The Bogdanov-Takens (BT), and
the saddle-node separatrix-loop (SNSL) codimension-two points.
The associated region of bistability between synchrony and asynchrony is bounded by}
homoclinic, saddle-node and Hopf bifurcations.

Remarkably, there is also a bistability region at small $\epsilon$ values for $r>r_*\simeq0.27891$
(recall the simulations in Fig.~\ref{fig:raster}),
which is bounded by a subcritical Hopf and a saddle-node of limit cycles bifurcations. 
In contrast to previous cases, this synchronization region is detached from the origin
due to the repulsive character of the iPRC. To be more precise, the bottom corner of the
lower bistability region located at point TC approaches the origin as $r\to1$.

\section{Conclusions}

In this work we have studied a non-trivial extension of the Winfree model in which
the PRC is nonlinear in the mean field. 
If the PRC contains only the first harmonic of the angle, the OA ansatz permits
a sharp dimensionality reduction.
Among all possible dependencies of the PRC on the mean field, we have considered only those 
with a crossover between two different canonical components. In particular, we have 
analyzed four cases
in which an attractive type-II component competes either against a repulsive type-II component 
or against a type-I component. Synchronization regions are peculiar for each case.
Bistability between macroscopic synchronization and
complete desynchronization are found in all cases (in case c, only for broad pulses),
but in different relative locations in the $\Delta-\epsilon$ plane.

Our results indicate
that the nonlinearity of the PRC with the forcing, by itself, is not enough to
generate complex collective phenomena. This is 
{\color{black} certain}
for a Lorentzian distribution of frequencies 
since the reduced system is only two dimensional, irrespective of the exact form of $f_1(A)$ and $f_2(A)$.
As happens in Kuramoto-like models, phenomena such as clustering or glassy dynamics may require 
multiple Fourier components \cite{okuda93} (in the PRC) or stronger heterogeneity \cite{IMS14}, respectively.
Concerning collective chaos, other ingredients such as
a time-varying coupling \cite{so11}, two interacting populations \cite{bick18} or 
multimodal frequency distributions \cite{cheng} appear to be imperative.

Needless to say, our study is only a drop in the ocean of possible PRCs and 
model generalizations. For instance, relaxation oscillators \cite{sacre16}
and bursting (neuronal) oscillators \cite{sherwood} have PRCs very different 
from the first-harmonic shape function in Eq.~\eqref{PRC}. 
Nevertheless, in spite of its limitations, we regard the
model defined by Eqs.~\eqref{model} and \eqref{PRC} as a noteworthy example of system in which the OA theory 
can be fully applied.

 \begin{acknowledgments}
We acknowledge support by the Agencia Estatal de Investigaci\'on and 
Fondo Europeo de Desarrollo Regional under Project No.~FIS2016-74957-P (AEI/FEDER, EU).
 \end{acknowledgments}

\section*{DATA AVAILABILITY}
 
 The data that support the findings of this study are available from the corresponding author 
 upon reasonable request.
 
\section*{Appendix: Identical Oscillators}

If the oscillators are identical there is a fully synchronized solution $\theta_j(t)=\Psi(t)$.
The dynamics of $\Psi$ obeys:
\[
\dot\Psi=\omega_0 + f_1[\varepsilon P(\Psi)](1-\cos\Psi)-f_2[\varepsilon
P(\Psi)]\sin\Psi.
\]
Next, we calculate the stability threshold of full synchrony, fixing $\omega_0=1$ as in the main text.
In the thermodynamic limit ($N\to\infty$), we may perturb one oscillator, say the first one, without changing the mean field.
Hence, one infinitesimal perturbation $\delta\theta=\theta_1-\Psi$ obeys:
\[
\dot{\delta\theta}=\lambda(\Psi)\, \delta\theta,
\]
where the multiplicative factor $\lambda(\Psi)=f_1[\varepsilon P(\Psi)]\sin\Psi-f_2[\varepsilon P(\Psi)]\cos\Psi$
depends on time through $\Psi(t)$.
In order to know the average exponential growth (or contraction) rate of $\delta\theta$ we need to integrate over variable
$\Psi$, taking into account its density $\rho(\Psi)$. These means that the sign
of constant $\lambda$, given by 
\[
 \lambda=\int_{-\pi}^{\pi} \lambda(\Psi) \rho(\Psi) d\Psi ,
\]
determines the stability of the fully synchronized solution. If $\lambda$ is positive, 
the oscillator ``evaporates'' from the main cluster, i.e. full synchrony is unstable.

The density $\rho(\Psi)$ is proportional to the inverse of the speed: $\rho(\Psi)\propto \dot\Psi^{-1}$.
Imposing $\lambda=0$, we obtain the condition for the stability threshold of full synchrony:
\[
\int_{-\pi}^{\pi} \frac{f_1[\varepsilon_c P(\Psi)]\sin\Psi-f_2[\varepsilon_c P(\Psi)]\cos\Psi}
{1 + f_1[\varepsilon_c P(\Psi)](1-\cos\Psi)-f_2[\varepsilon_c P(\Psi)]\sin\Psi}
d\Psi=0.
\]
This integral cannot be solved analytically, but the threshold coupling $\varepsilon_c$ is easily found
numerically.

\section*{References}


\begin{thebibliography}{48}%
\makeatletter
\providecommand \@ifxundefined [1]{%
 \@ifx{#1\undefined}
}%
\providecommand \@ifnum [1]{%
 \ifnum #1\expandafter \@firstoftwo
 \else \expandafter \@secondoftwo
 \fi
}%
\providecommand \@ifx [1]{%
 \ifx #1\expandafter \@firstoftwo
 \else \expandafter \@secondoftwo
 \fi
}%
\providecommand \natexlab [1]{#1}%
\providecommand \enquote  [1]{``#1''}%
\providecommand \bibnamefont  [1]{#1}%
\providecommand \bibfnamefont [1]{#1}%
\providecommand \citenamefont [1]{#1}%
\providecommand \href@noop [0]{\@secondoftwo}%
\providecommand \href [0]{\begingroup \@sanitize@url \@href}%
\providecommand \@href[1]{\@@startlink{#1}\@@href}%
\providecommand \@@href[1]{\endgroup#1\@@endlink}%
\providecommand \@sanitize@url [0]{\catcode `\\12\catcode `\$12\catcode
  `\&12\catcode `\#12\catcode `\^12\catcode `\_12\catcode `\%12\relax}%
\providecommand \@@startlink[1]{}%
\providecommand \@@endlink[0]{}%
\providecommand \url  [0]{\begingroup\@sanitize@url \@url }%
\providecommand \@url [1]{\endgroup\@href {#1}{\urlprefix }}%
\providecommand \urlprefix  [0]{URL }%
\providecommand \Eprint [0]{\href }%
\providecommand \doibase [0]{http://dx.doi.org/}%
\providecommand \selectlanguage [0]{\@gobble}%
\providecommand \bibinfo  [0]{\@secondoftwo}%
\providecommand \bibfield  [0]{\@secondoftwo}%
\providecommand \translation [1]{[#1]}%
\providecommand \BibitemOpen [0]{}%
\providecommand \bibitemStop [0]{}%
\providecommand \bibitemNoStop [0]{.\EOS\space}%
\providecommand \EOS [0]{\spacefactor3000\relax}%
\providecommand \BibitemShut  [1]{\csname bibitem#1\endcsname}%
\let\auto@bib@innerbib\@empty
\bibitem [{\citenamefont {Winfree}(1967)}]{Win67}%
  \BibitemOpen
  \bibfield  {author} {\bibinfo {author} {\bibfnamefont {A.~T.}\ \bibnamefont
  {Winfree}},\ }\bibfield  {title} {\enquote {\bibinfo {title} {Biological
  rhythms and the behavior of populations of coupled oscillators.}}\
  }\href@noop {} {\bibfield  {journal} {\bibinfo  {journal} {J. Theor. Biol.}\
  }\textbf {\bibinfo {volume} {16}},\ \bibinfo {pages} {15--42} (\bibinfo
  {year} {1967})}\BibitemShut {NoStop}%
\bibitem [{\citenamefont {Kuramoto}(1975)}]{Kur75}%
  \BibitemOpen
  \bibfield  {author} {\bibinfo {author} {\bibfnamefont {Y.}~\bibnamefont
  {Kuramoto}},\ }\bibfield  {title} {\enquote {\bibinfo {title}
  {Self-entrainment of a population of coupled non-linear oscillators},}\ }in\
  \href@noop {} {\emph {\bibinfo {booktitle} {International Symposium on
  Mathematical Problems in Theoretical Physics}}},\ \bibinfo {series} {Lecture
  Notes in Physics}, Vol.~\bibinfo {volume} {39},\ \bibinfo {editor} {edited
  by\ \bibinfo {editor} {\bibfnamefont {H.}~\bibnamefont {Araki}}}\ (\bibinfo
  {publisher} {Springer},\ \bibinfo {address} {Berlin},\ \bibinfo {year}
  {1975})\ pp.\ \bibinfo {pages} {420--422}\BibitemShut {NoStop}%
\bibitem [{\citenamefont {Strogatz}(2000)}]{Str00}%
  \BibitemOpen
  \bibfield  {author} {\bibinfo {author} {\bibfnamefont {S.~H.}\ \bibnamefont
  {Strogatz}},\ }\bibfield  {title} {\enquote {\bibinfo {title} {From
  {K}uramoto to {C}rawford: exploring the onset of synchronization in
  populations of coupled oscillators},}\ }\href {\doibase
  10.1016/S0167-2789(00)00094-4} {\bibfield  {journal} {\bibinfo  {journal}
  {Physica D}\ }\textbf {\bibinfo {volume} {143}},\ \bibinfo {pages} {1--20}
  (\bibinfo {year} {2000})}\BibitemShut {NoStop}%
\bibitem [{\citenamefont {Montbri\'o}\ and\ \citenamefont
  {Paz\'o}(2018)}]{MP18}%
  \BibitemOpen
  \bibfield  {author} {\bibinfo {author} {\bibfnamefont {E.}~\bibnamefont
  {Montbri\'o}}\ and\ \bibinfo {author} {\bibfnamefont {D.}~\bibnamefont
  {Paz\'o}},\ }\bibfield  {title} {\enquote {\bibinfo {title} {Kuramoto model
  for excitation-inhibition-based oscillations},}\ }\href {\doibase
  10.1103/PhysRevLett.120.244101} {\bibfield  {journal} {\bibinfo  {journal}
  {Phys. Rev. Lett.}\ }\textbf {\bibinfo {volume} {120}},\ \bibinfo {pages}
  {244101} (\bibinfo {year} {2018})}\BibitemShut {NoStop}%
\bibitem [{\citenamefont {Paz\'o}\ and\ \citenamefont
  {Montbri\'o}(2014)}]{PM14}%
  \BibitemOpen
  \bibfield  {author} {\bibinfo {author} {\bibfnamefont {D.}~\bibnamefont
  {Paz\'o}}\ and\ \bibinfo {author} {\bibfnamefont {E.}~\bibnamefont
  {Montbri\'o}},\ }\bibfield  {title} {\enquote {\bibinfo {title}
  {Low-dimensional dynamics of populations of pulse-coupled oscillators},}\
  }\href {\doibase 10.1103/PhysRevX.4.011009} {\bibfield  {journal} {\bibinfo
  {journal} {Phys. Rev. X}\ }\textbf {\bibinfo {volume} {4}},\ \bibinfo {pages}
  {011009} (\bibinfo {year} {2014})}\BibitemShut {NoStop}%
\bibitem [{\citenamefont {Gallego}, \citenamefont {Montbri\'o},\ and\
  \citenamefont {Paz\'o}(2017)}]{gallego17}%
  \BibitemOpen
  \bibfield  {author} {\bibinfo {author} {\bibfnamefont {R.}~\bibnamefont
  {Gallego}}, \bibinfo {author} {\bibfnamefont {E.}~\bibnamefont {Montbri\'o}},
  \ and\ \bibinfo {author} {\bibfnamefont {D.}~\bibnamefont {Paz\'o}},\
  }\bibfield  {title} {\enquote {\bibinfo {title} {Synchronization scenarios in
  the {W}infree model of coupled oscillators},}\ }\href {\doibase
  10.1103/PhysRevE.96.042208} {\bibfield  {journal} {\bibinfo  {journal} {Phys.
  Rev. E}\ }\textbf {\bibinfo {volume} {96}},\ \bibinfo {pages} {042208}
  (\bibinfo {year} {2017})}\BibitemShut {NoStop}%
\bibitem [{\citenamefont {Ariaratnam}\ and\ \citenamefont
  {Strogatz}(2001)}]{AS01}%
  \BibitemOpen
  \bibfield  {author} {\bibinfo {author} {\bibfnamefont {J.~T.}\ \bibnamefont
  {Ariaratnam}}\ and\ \bibinfo {author} {\bibfnamefont {S.~H.}\ \bibnamefont
  {Strogatz}},\ }\bibfield  {title} {\enquote {\bibinfo {title} {Phase diagram
  for the {W}infree model of coupled nonlinear oscillators},}\ }\href {\doibase
  10.1103/PhysRevLett.86.4278} {\bibfield  {journal} {\bibinfo  {journal}
  {Phys. Rev. Lett.}\ }\textbf {\bibinfo {volume} {86}},\ \bibinfo {pages}
  {4278--4281} (\bibinfo {year} {2001})}\BibitemShut {NoStop}%
\bibitem [{\citenamefont {Ott}\ and\ \citenamefont {Antonsen}(2008)}]{OA08}%
  \BibitemOpen
  \bibfield  {author} {\bibinfo {author} {\bibfnamefont {E.}~\bibnamefont
  {Ott}}\ and\ \bibinfo {author} {\bibfnamefont {T.~M.}\ \bibnamefont
  {Antonsen}},\ }\bibfield  {title} {\enquote {\bibinfo {title} {Low
  dimensional behavior of large systems of globally coupled oscillators},}\
  }\href {\doibase 10.1063/1.2930766} {\bibfield  {journal} {\bibinfo
  {journal} {Chaos}\ }\textbf {\bibinfo {volume} {18}},\ \bibinfo {eid}
  {037113} (\bibinfo {year} {2008})}\BibitemShut {NoStop}%
\bibitem [{\citenamefont {Ott}\ and\ \citenamefont {Antonsen}(2009)}]{OA09}%
  \BibitemOpen
  \bibfield  {author} {\bibinfo {author} {\bibfnamefont {E.}~\bibnamefont
  {Ott}}\ and\ \bibinfo {author} {\bibfnamefont {T.~M.}\ \bibnamefont
  {Antonsen}},\ }\bibfield  {title} {\enquote {\bibinfo {title} {Long time
  evolution of phase oscillator systems},}\ }\href {\doibase 10.1063/1.3136851}
  {\bibfield  {journal} {\bibinfo  {journal} {Chaos}\ }\textbf {\bibinfo
  {volume} {19}},\ \bibinfo {eid} {023117} (\bibinfo {year}
  {2009})}\BibitemShut {NoStop}%
\bibitem [{\citenamefont {Ott}, \citenamefont {Hunt},\ and\ \citenamefont
  {Antonsen}(2011)}]{OHA11}%
  \BibitemOpen
  \bibfield  {author} {\bibinfo {author} {\bibfnamefont {E.}~\bibnamefont
  {Ott}}, \bibinfo {author} {\bibfnamefont {B.~R.}\ \bibnamefont {Hunt}}, \
  and\ \bibinfo {author} {\bibfnamefont {T.~M.}\ \bibnamefont {Antonsen}},\
  }\bibfield  {title} {\enquote {\bibinfo {title} {Comment on ``long time
  evolution of phase oscillators systems''},}\ }\href {\doibase
  10.1063/1.3574931} {\bibfield  {journal} {\bibinfo  {journal} {Chaos}\
  }\textbf {\bibinfo {volume} {21}},\ \bibinfo {eid} {025112} (\bibinfo {year}
  {2011})}\BibitemShut {NoStop}%
\bibitem [{\citenamefont {Winfree}(1980)}]{Win80}%
  \BibitemOpen
  \bibfield  {author} {\bibinfo {author} {\bibfnamefont {A.~T.}\ \bibnamefont
  {Winfree}},\ }\href@noop {} {\emph {\bibinfo {title} {The Geometry of
  Biological Time}}}\ (\bibinfo  {publisher} {Springer},\ \bibinfo {address}
  {New York},\ \bibinfo {year} {1980})\BibitemShut {NoStop}%
\bibitem [{\citenamefont {Pikovsky}, \citenamefont {Rosenblum},\ and\
  \citenamefont {Kurths}(2001)}]{PRK01}%
  \BibitemOpen
  \bibfield  {author} {\bibinfo {author} {\bibfnamefont {A.~S.}\ \bibnamefont
  {Pikovsky}}, \bibinfo {author} {\bibfnamefont {M.~G.}\ \bibnamefont
  {Rosenblum}}, \ and\ \bibinfo {author} {\bibfnamefont {J.}~\bibnamefont
  {Kurths}},\ }\href@noop {} {\emph {\bibinfo {title} {Synchronization, a
  Universal Concept in Nonlinear Sciences}}}\ (\bibinfo  {publisher} {Cambridge
  University Press},\ \bibinfo {address} {Cambridge},\ \bibinfo {year}
  {2001})\BibitemShut {NoStop}%
\bibitem [{\citenamefont {Izhikevich}(2007)}]{Izh07}%
  \BibitemOpen
  \bibfield  {author} {\bibinfo {author} {\bibfnamefont {E.~M.}\ \bibnamefont
  {Izhikevich}},\ }\href@noop {} {\emph {\bibinfo {title} {Dynamical Systems in
  Neuroscience}}}\ (\bibinfo  {publisher} {The MIT Press},\ \bibinfo {address}
  {Cambridge, Massachusetts},\ \bibinfo {year} {2007})\ Chap.~\bibinfo
  {chapter} {10}\BibitemShut {NoStop}%
\bibitem [{\citenamefont {Kuramoto}(1984)}]{Kur84}%
  \BibitemOpen
  \bibfield  {author} {\bibinfo {author} {\bibfnamefont {Y.}~\bibnamefont
  {Kuramoto}},\ }\href@noop {} {\emph {\bibinfo {title} {Chemical Oscillations,
  Waves, and Turbulence}}}\ (\bibinfo  {publisher} {{S}pringer-{V}erlag},\
  \bibinfo {address} {Berlin},\ \bibinfo {year} {1984})\BibitemShut {NoStop}%
\bibitem [{\citenamefont {Sacr{\'e}}\ and\ \citenamefont
  {{Sepulchre}}(2014)}]{sacre14}%
  \BibitemOpen
  \bibfield  {author} {\bibinfo {author} {\bibfnamefont {P.}~\bibnamefont
  {Sacr{\'e}}}\ and\ \bibinfo {author} {\bibfnamefont {R.}~\bibnamefont
  {{Sepulchre}}},\ }\bibfield  {title} {\enquote {\bibinfo {title} {Sensitivity
  analysis of oscillator models in the space of phase-response curves:
  Oscillators as open systems},}\ }\href {\doibase 10.1109/MCS.2013.2295710}
  {\bibfield  {journal} {\bibinfo  {journal} {IEEE Control Systems Magazine}\
  }\textbf {\bibinfo {volume} {34}},\ \bibinfo {pages} {50--74} (\bibinfo
  {year} {2014})}\BibitemShut {NoStop}%
\bibitem [{\citenamefont {Pietras}\ and\ \citenamefont
  {Daffertshofer}(2019)}]{pietras19}%
  \BibitemOpen
  \bibfield  {author} {\bibinfo {author} {\bibfnamefont {B.}~\bibnamefont
  {Pietras}}\ and\ \bibinfo {author} {\bibfnamefont {A.}~\bibnamefont
  {Daffertshofer}},\ }\bibfield  {title} {\enquote {\bibinfo {title} {Network
  dynamics of coupled oscillators and phase reduction techniques},}\ }\href
  {\doibase https://doi.org/10.1016/j.physrep.2019.06.001} {\bibfield
  {journal} {\bibinfo  {journal} {Phys. Rep.}\ }\textbf {\bibinfo {volume}
  {819}},\ \bibinfo {pages} {1 -- 105} (\bibinfo {year} {2019})},\ \bibinfo
  {note} {network dynamics of coupled oscillators and phase reduction
  techniques}\BibitemShut {NoStop}%
\bibitem [{\citenamefont {Ermentrout}\ and\ \citenamefont
  {Terman}(2010)}]{ET10}%
  \BibitemOpen
  \bibfield  {author} {\bibinfo {author} {\bibfnamefont {G.~B.}\ \bibnamefont
  {Ermentrout}}\ and\ \bibinfo {author} {\bibfnamefont {D.~H.}\ \bibnamefont
  {Terman}},\ }\href@noop {} {\emph {\bibinfo {title} {Mathematical Foundations
  of Neuroscience}}},\ Vol.~\bibinfo {volume} {64}\ (\bibinfo  {publisher}
  {Springer},\ \bibinfo {year} {2010})\BibitemShut {NoStop}%
\bibitem [{\citenamefont {B{\"o}rgers}(2017)}]{Boergers}%
  \BibitemOpen
  \bibfield  {author} {\bibinfo {author} {\bibfnamefont {C.}~\bibnamefont
  {B{\"o}rgers}},\ }\href {https://books.google.es/books?id=YLuwDgAAQBAJ}
  {\emph {\bibinfo {title} {An Introduction to Modeling Neuronal Dynamics}}},\
  Texts in Applied Mathematics\ (\bibinfo  {publisher} {Springer International
  Publishing},\ \bibinfo {year} {2017})\BibitemShut {NoStop}%
\bibitem [{\citenamefont {Rode}\ \emph {et~al.}(2019)\citenamefont {Rode},
  \citenamefont {Totz}, \citenamefont {Fengler},\ and\ \citenamefont
  {Engel}}]{rode19}%
  \BibitemOpen
  \bibfield  {author} {\bibinfo {author} {\bibfnamefont {J.}~\bibnamefont
  {Rode}}, \bibinfo {author} {\bibfnamefont {J.~F.}\ \bibnamefont {Totz}},
  \bibinfo {author} {\bibfnamefont {E.}~\bibnamefont {Fengler}}, \ and\
  \bibinfo {author} {\bibfnamefont {H.}~\bibnamefont {Engel}},\ }\bibfield
  {title} {\enquote {\bibinfo {title} {Chimera states on a ring of strongly
  coupled relaxation oscillators},}\ }\href {\doibase 10.3389/fams.2019.00031}
  {\bibfield  {journal} {\bibinfo  {journal} {Frontiers in Applied Mathematics
  and Statistics}\ }\textbf {\bibinfo {volume} {5}},\ \bibinfo {pages} {31}
  (\bibinfo {year} {2019})}\BibitemShut {NoStop}%
\bibitem [{\citenamefont {C\u{a}lug\u{a}ru}\ \emph {et~al.}()\citenamefont
  {C\u{a}lug\u{a}ru}, \citenamefont {Totz}, \citenamefont {Martens},\ and\
  \citenamefont {Engel}}]{totz}%
  \BibitemOpen
  \bibfield  {author} {\bibinfo {author} {\bibfnamefont {D.}~\bibnamefont
  {C\u{a}lug\u{a}ru}}, \bibinfo {author} {\bibfnamefont {J.~F.}\ \bibnamefont
  {Totz}}, \bibinfo {author} {\bibfnamefont {E.~A.}\ \bibnamefont {Martens}}, \
  and\ \bibinfo {author} {\bibfnamefont {H.}~\bibnamefont {Engel}},\
  }\href@noop {} {\enquote {\bibinfo {title} {First-order synchronization
  transition in a large population of relaxation oscillators},}\ }\bibinfo
  {howpublished} {arXiv:1812.04727}\BibitemShut {NoStop}%
\bibitem [{\citenamefont {Paz{\'{o}}}, \citenamefont {Montbri{\'{o}}},\ and\
  \citenamefont {Gallego}(2019)}]{PMG19}%
  \BibitemOpen
  \bibfield  {author} {\bibinfo {author} {\bibfnamefont {D.}~\bibnamefont
  {Paz{\'{o}}}}, \bibinfo {author} {\bibfnamefont {E.}~\bibnamefont
  {Montbri{\'{o}}}}, \ and\ \bibinfo {author} {\bibfnamefont {R.}~\bibnamefont
  {Gallego}},\ }\bibfield  {title} {\enquote {\bibinfo {title} {The {W}infree
  model with heterogeneous phase-response curves: analytical results},}\ }\href
  {\doibase 10.1088/1751-8121/ab0b4c} {\bibfield  {journal} {\bibinfo
  {journal} {J. Phys. A: Math. and Theor.}\ }\textbf {\bibinfo {volume} {52}},\
  \bibinfo {pages} {154001} (\bibinfo {year} {2019})}\BibitemShut {NoStop}%
\bibitem [{\citenamefont {Luke}, \citenamefont {Barreto},\ and\ \citenamefont
  {So}(2013)}]{LBS13}%
  \BibitemOpen
  \bibfield  {author} {\bibinfo {author} {\bibfnamefont {T.~B.}\ \bibnamefont
  {Luke}}, \bibinfo {author} {\bibfnamefont {E.}~\bibnamefont {Barreto}}, \
  and\ \bibinfo {author} {\bibfnamefont {P.}~\bibnamefont {So}},\ }\bibfield
  {title} {\enquote {\bibinfo {title} {Complete classification of the
  macroscopic behavior of a heterogeneous network of theta neurons},}\
  }\href@noop {} {\bibfield  {journal} {\bibinfo  {journal} {Neural Comput.}\
  }\textbf {\bibinfo {volume} {25}},\ \bibinfo {pages} {3207--3234} (\bibinfo
  {year} {2013})}\BibitemShut {NoStop}%
\bibitem [{\citenamefont {Laing}(2014)}]{laing14}%
  \BibitemOpen
  \bibfield  {author} {\bibinfo {author} {\bibfnamefont {C.~R.}\ \bibnamefont
  {Laing}},\ }\bibfield  {title} {\enquote {\bibinfo {title} {Derivation of a
  neural field model from a network of theta neurons},}\ }\href {\doibase
  10.1103/PhysRevE.90.010901} {\bibfield  {journal} {\bibinfo  {journal} {Phys.
  Rev. E}\ }\textbf {\bibinfo {volume} {90}},\ \bibinfo {pages} {010901}
  (\bibinfo {year} {2014})}\BibitemShut {NoStop}%
\bibitem [{\citenamefont {So}, \citenamefont {Luke},\ and\ \citenamefont
  {Barreto}(2014)}]{SLB14}%
  \BibitemOpen
  \bibfield  {author} {\bibinfo {author} {\bibfnamefont {P.}~\bibnamefont
  {So}}, \bibinfo {author} {\bibfnamefont {T.~B.}\ \bibnamefont {Luke}}, \ and\
  \bibinfo {author} {\bibfnamefont {E.}~\bibnamefont {Barreto}},\ }\bibfield
  {title} {\enquote {\bibinfo {title} {Networks of theta neurons with
  time-varying excitability: Macroscopic chaos, multistability, and final-state
  uncertainty},}\ }\href {\doibase
  http://dx.doi.org/10.1016/j.physd.2013.04.009} {\bibfield  {journal}
  {\bibinfo  {journal} {Physica D}\ }\textbf {\bibinfo {volume} {267}},\
  \bibinfo {pages} {16--26} (\bibinfo {year} {2014})}\BibitemShut {NoStop}%
\bibitem [{\citenamefont {Montbri\'o}, \citenamefont {Paz\'o},\ and\
  \citenamefont {Roxin}(2015)}]{MPR15}%
  \BibitemOpen
  \bibfield  {author} {\bibinfo {author} {\bibfnamefont {E.}~\bibnamefont
  {Montbri\'o}}, \bibinfo {author} {\bibfnamefont {D.}~\bibnamefont {Paz\'o}},
  \ and\ \bibinfo {author} {\bibfnamefont {A.}~\bibnamefont {Roxin}},\
  }\bibfield  {title} {\enquote {\bibinfo {title} {Macroscopic description for
  networks of spiking neurons},}\ }\href {\doibase 10.1103/PhysRevX.5.021028}
  {\bibfield  {journal} {\bibinfo  {journal} {Phys. Rev. X}\ }\textbf {\bibinfo
  {volume} {5}},\ \bibinfo {pages} {021028} (\bibinfo {year}
  {2015})}\BibitemShut {NoStop}%
\bibitem [{\citenamefont {Paz\'o}\ and\ \citenamefont
  {Montbri\'o}(2016)}]{PM16}%
  \BibitemOpen
  \bibfield  {author} {\bibinfo {author} {\bibfnamefont {D.}~\bibnamefont
  {Paz\'o}}\ and\ \bibinfo {author} {\bibfnamefont {E.}~\bibnamefont
  {Montbri\'o}},\ }\bibfield  {title} {\enquote {\bibinfo {title} {From
  quasiperiodic partial synchronization to collective chaos in populations of
  inhibitory neurons with delay},}\ }\href {\doibase
  10.1103/PhysRevLett.116.238101} {\bibfield  {journal} {\bibinfo  {journal}
  {Phys. Rev. Lett.}\ }\textbf {\bibinfo {volume} {116}},\ \bibinfo {pages}
  {238101} (\bibinfo {year} {2016})}\BibitemShut {NoStop}%
\bibitem [{\citenamefont {Ratas}\ and\ \citenamefont {Pyragas}(2016)}]{RP16}%
  \BibitemOpen
  \bibfield  {author} {\bibinfo {author} {\bibfnamefont {I.}~\bibnamefont
  {Ratas}}\ and\ \bibinfo {author} {\bibfnamefont {K.}~\bibnamefont
  {Pyragas}},\ }\bibfield  {title} {\enquote {\bibinfo {title} {Macroscopic
  self-oscillations and aging transition in a network of synaptically coupled
  quadratic integrate-and-fire neurons},}\ }\href {\doibase
  10.1103/PhysRevE.94.032215} {\bibfield  {journal} {\bibinfo  {journal} {Phys.
  Rev. E}\ }\textbf {\bibinfo {volume} {94}},\ \bibinfo {pages} {032215}
  (\bibinfo {year} {2016})}\BibitemShut {NoStop}%
\bibitem [{\citenamefont {O'Keeffe}\ and\ \citenamefont
  {Strogatz}(2016)}]{okeeffe16}%
  \BibitemOpen
  \bibfield  {author} {\bibinfo {author} {\bibfnamefont {K.~P.}\ \bibnamefont
  {O'Keeffe}}\ and\ \bibinfo {author} {\bibfnamefont {S.~H.}\ \bibnamefont
  {Strogatz}},\ }\bibfield  {title} {\enquote {\bibinfo {title} {Dynamics of a
  population of oscillatory and excitable elements},}\ }\href {\doibase
  10.1103/PhysRevE.93.062203} {\bibfield  {journal} {\bibinfo  {journal} {Phys.
  Rev. E}\ }\textbf {\bibinfo {volume} {93}},\ \bibinfo {pages} {062203}
  (\bibinfo {year} {2016})}\BibitemShut {NoStop}%
\bibitem [{\citenamefont {Roulet}\ and\ \citenamefont
  {Mindlin}(2016)}]{roulet16}%
  \BibitemOpen
  \bibfield  {author} {\bibinfo {author} {\bibfnamefont {J.}~\bibnamefont
  {Roulet}}\ and\ \bibinfo {author} {\bibfnamefont {G.~B.}\ \bibnamefont
  {Mindlin}},\ }\bibfield  {title} {\enquote {\bibinfo {title} {Average
  activity of excitatory and inhibitory neural populations},}\ }\href {\doibase
  10.1063/1.4962326} {\bibfield  {journal} {\bibinfo  {journal} {Chaos}\
  }\textbf {\bibinfo {volume} {26}},\ \bibinfo {pages} {093104} (\bibinfo
  {year} {2016})}\BibitemShut {NoStop}%
\bibitem [{\citenamefont {Hansel}, \citenamefont {Mato},\ and\ \citenamefont
  {Meunier}(1995)}]{hansel95}%
  \BibitemOpen
  \bibfield  {author} {\bibinfo {author} {\bibfnamefont {D.}~\bibnamefont
  {Hansel}}, \bibinfo {author} {\bibfnamefont {G.}~\bibnamefont {Mato}}, \ and\
  \bibinfo {author} {\bibfnamefont {C.}~\bibnamefont {Meunier}},\ }\bibfield
  {title} {\enquote {\bibinfo {title} {Synchrony in excitatory neural
  networks},}\ }\href@noop {} {\bibfield  {journal} {\bibinfo  {journal}
  {Neural Comput.}\ }\textbf {\bibinfo {volume} {7}},\ \bibinfo {pages}
  {307--337} (\bibinfo {year} {1995})}\BibitemShut {NoStop}%
\bibitem [{\citenamefont {Reyes}\ and\ \citenamefont {Fetz}(1993)}]{reyes93}%
  \BibitemOpen
  \bibfield  {author} {\bibinfo {author} {\bibfnamefont {A.~D.}\ \bibnamefont
  {Reyes}}\ and\ \bibinfo {author} {\bibfnamefont {E.~E.}\ \bibnamefont
  {Fetz}},\ }\bibfield  {title} {\enquote {\bibinfo {title} {Two modes of
  interspike interval shortening by brief transient depolarizations in cat
  neocortical neurons},}\ }\href {\doibase 10.1152/jn.1993.69.5.1661}
  {\bibfield  {journal} {\bibinfo  {journal} {J. Neurophysiol.}\ }\textbf
  {\bibinfo {volume} {69}},\ \bibinfo {pages} {1661--1672} (\bibinfo {year}
  {1993})}\BibitemShut {NoStop}%
\bibitem [{\citenamefont {Netoff}\ \emph {et~al.}(2005)\citenamefont {Netoff},
  \citenamefont {Banks}, \citenamefont {Dorval}, \citenamefont {Acker},
  \citenamefont {Haas}, \citenamefont {Kopell},\ and\ \citenamefont
  {White}}]{netoff05}%
  \BibitemOpen
  \bibfield  {author} {\bibinfo {author} {\bibfnamefont {T.~I.}\ \bibnamefont
  {Netoff}}, \bibinfo {author} {\bibfnamefont {M.~I.}\ \bibnamefont {Banks}},
  \bibinfo {author} {\bibfnamefont {A.~D.}\ \bibnamefont {Dorval}}, \bibinfo
  {author} {\bibfnamefont {C.~D.}\ \bibnamefont {Acker}}, \bibinfo {author}
  {\bibfnamefont {J.~S.}\ \bibnamefont {Haas}}, \bibinfo {author}
  {\bibfnamefont {N.}~\bibnamefont {Kopell}}, \ and\ \bibinfo {author}
  {\bibfnamefont {J.~A.}\ \bibnamefont {White}},\ }\bibfield  {title} {\enquote
  {\bibinfo {title} {Synchronization in hybrid neuronal networks of the
  hippocampal formation},}\ }\href {\doibase 10.1152/jn.00982.2004} {\bibfield
  {journal} {\bibinfo  {journal} {J. Neurophysiol.}\ }\textbf {\bibinfo
  {volume} {93}},\ \bibinfo {pages} {1197--1208} (\bibinfo {year}
  {2005})}\BibitemShut {NoStop}%
\bibitem [{\citenamefont {Buck}(1988)}]{buck88}%
  \BibitemOpen
  \bibfield  {author} {\bibinfo {author} {\bibfnamefont {J.}~\bibnamefont
  {Buck}},\ }\bibfield  {title} {\enquote {\bibinfo {title} {Synchronous
  rhythmic flashing of fireflies {II}},}\ }\href {\doibase 10.1086/415929}
  {\bibfield  {journal} {\bibinfo  {journal} {Q. Rev. Biol.}\ }\textbf
  {\bibinfo {volume} {63}},\ \bibinfo {pages} {265--289} (\bibinfo {year}
  {1988})}\BibitemShut {NoStop}%
\bibitem [{\citenamefont {Hanson}(1978)}]{hanson78}%
  \BibitemOpen
  \bibfield  {author} {\bibinfo {author} {\bibfnamefont {F.~E.}\ \bibnamefont
  {Hanson}},\ }\bibfield  {title} {\enquote {\bibinfo {title} {Comparative
  studies of firefly pacemakers},}\ }\href@noop {} {\bibfield  {journal}
  {\bibinfo  {journal} {Fed. Proc.}\ }\textbf {\bibinfo {volume} {37}},\
  \bibinfo {pages} {2158--2164} (\bibinfo {year} {1978})}\BibitemShut {NoStop}%
\bibitem [{\citenamefont {Pietras}\ and\ \citenamefont
  {Daffertshofer}(2016)}]{PD16}%
  \BibitemOpen
  \bibfield  {author} {\bibinfo {author} {\bibfnamefont {B.}~\bibnamefont
  {Pietras}}\ and\ \bibinfo {author} {\bibfnamefont {A.}~\bibnamefont
  {Daffertshofer}},\ }\bibfield  {title} {\enquote {\bibinfo {title}
  {Ott-{A}ntonsen attractiveness for parameter-dependent oscillatory
  systems},}\ }\href {\doibase 10.1063/1.4963371} {\bibfield  {journal}
  {\bibinfo  {journal} {Chaos}\ }\textbf {\bibinfo {volume} {26}},\ \bibinfo
  {pages} {103101} (\bibinfo {year} {2016})}\BibitemShut {NoStop}%
\bibitem [{\citenamefont {Dhooge}, \citenamefont {Govaerts},\ and\
  \citenamefont {Kuznetsov}(2004)}]{matcont}%
  \BibitemOpen
  \bibfield  {author} {\bibinfo {author} {\bibfnamefont {A.}~\bibnamefont
  {Dhooge}}, \bibinfo {author} {\bibfnamefont {W.}~\bibnamefont {Govaerts}}, \
  and\ \bibinfo {author} {\bibfnamefont {Y.~A.}\ \bibnamefont {Kuznetsov}},\
  }\bibfield  {title} {\enquote {\bibinfo {title} {{MATCONT}: A matlab package
  for numerical bifurcation analysis of {ODE}s},}\ }\href {\doibase
  10.1145/980175.980184} {\bibfield  {journal} {\bibinfo  {journal} {SIGSAM
  Bull.}\ }\textbf {\bibinfo {volume} {38}},\ \bibinfo {pages} {21–22}
  (\bibinfo {year} {2004})}\BibitemShut {NoStop}%
\bibitem [{\citenamefont {Ashwin}\ and\ \citenamefont
  {Swift}(1992)}]{ashwin92}%
  \BibitemOpen
  \bibfield  {author} {\bibinfo {author} {\bibfnamefont {P.}~\bibnamefont
  {Ashwin}}\ and\ \bibinfo {author} {\bibfnamefont {J.~W.}\ \bibnamefont
  {Swift}},\ }\bibfield  {title} {\enquote {\bibinfo {title} {The dynamics of n
  weakly coupled identical oscillators},}\ }\href {\doibase 10.1007/BF02429852}
  {\bibfield  {journal} {\bibinfo  {journal} {J. Nonlin. Sci.}\ }\textbf
  {\bibinfo {volume} {2}},\ \bibinfo {pages} {69--108} (\bibinfo {year}
  {1992})}\BibitemShut {NoStop}%
\bibitem [{Note1()}]{Note1}%
  \BibitemOpen
  \bibinfo {note} {As the ``OA manifold'' is not attracting for identical
  oscillators \cite {OA08,OA09,OHA11}, the resulting dynamics depends on the
  initial conditions. (For initial conditions in the OA manifold, such as
  purely random phases ($Z=0$), the system converges to a state of
  quasiperiodic partial synchronization \cite {Vre96,PR15}, a state which $Z$
  oscillates periodically while the individual oscillators behave
  quasiperiodically.)}\BibitemShut {NoStop}%
\bibitem [{\citenamefont {Crawford}(1991)}]{crawford91}%
  \BibitemOpen
  \bibfield  {author} {\bibinfo {author} {\bibfnamefont {J.~D.}\ \bibnamefont
  {Crawford}},\ }\bibfield  {title} {\enquote {\bibinfo {title} {Introduction
  to bifurcation theory},}\ }\href {\doibase 10.1103/RevModPhys.63.991}
  {\bibfield  {journal} {\bibinfo  {journal} {Rev. Mod. Phys.}\ }\textbf
  {\bibinfo {volume} {63}},\ \bibinfo {pages} {991--1037} (\bibinfo {year}
  {1991})}\BibitemShut {NoStop}%
\bibitem [{\citenamefont {Okuda}(1993)}]{okuda93}%
  \BibitemOpen
  \bibfield  {author} {\bibinfo {author} {\bibfnamefont {K.}~\bibnamefont
  {Okuda}},\ }\bibfield  {title} {\enquote {\bibinfo {title} {Variety and
  generality of clustering in globally coupled oscillators},}\ }\href {\doibase
  https://doi.org/10.1016/0167-2789(93)90121-G} {\bibfield  {journal} {\bibinfo
   {journal} {Physica D: Nonlinear Phenomena}\ }\textbf {\bibinfo {volume}
  {63}},\ \bibinfo {pages} {424 -- 436} (\bibinfo {year} {1993})}\BibitemShut
  {NoStop}%
\bibitem [{\citenamefont {Iatsenko}, \citenamefont {McClintock},\ and\
  \citenamefont {Stefanovska}(2014)}]{IMS14}%
  \BibitemOpen
  \bibfield  {author} {\bibinfo {author} {\bibfnamefont {D.}~\bibnamefont
  {Iatsenko}}, \bibinfo {author} {\bibfnamefont {P.~V.~E.}\ \bibnamefont
  {McClintock}}, \ and\ \bibinfo {author} {\bibfnamefont {A.}~\bibnamefont
  {Stefanovska}},\ }\bibfield  {title} {\enquote {\bibinfo {title} {Oscillator
  glass in the generalized {K}uramoto model: synchronous disorder and two-step
  relaxation},}\ }\href {\doibase 10.1038/ncomms5118} {\bibfield  {journal}
  {\bibinfo  {journal} {Nat. Commun.}\ }\textbf {\bibinfo {volume} {5}},\
  \bibinfo {pages} {4188} (\bibinfo {year} {2014})}\BibitemShut {NoStop}%
\bibitem [{\citenamefont {So}\ and\ \citenamefont {Barreto}(2011)}]{so11}%
  \BibitemOpen
  \bibfield  {author} {\bibinfo {author} {\bibfnamefont {P.}~\bibnamefont
  {So}}\ and\ \bibinfo {author} {\bibfnamefont {E.}~\bibnamefont {Barreto}},\
  }\bibfield  {title} {\enquote {\bibinfo {title} {Generating macroscopic chaos
  in a network of globally coupled phase oscillators},}\ }\href {\doibase
  10.1063/1.3638441} {\bibfield  {journal} {\bibinfo  {journal} {Chaos}\
  }\textbf {\bibinfo {volume} {21}},\ \bibinfo {pages} {033127} (\bibinfo
  {year} {2011})}\BibitemShut {NoStop}%
\bibitem [{\citenamefont {Bick}, \citenamefont {Panaggio},\ and\ \citenamefont
  {Martens}(2018)}]{bick18}%
  \BibitemOpen
  \bibfield  {author} {\bibinfo {author} {\bibfnamefont {C.}~\bibnamefont
  {Bick}}, \bibinfo {author} {\bibfnamefont {M.~J.}\ \bibnamefont {Panaggio}},
  \ and\ \bibinfo {author} {\bibfnamefont {E.~A.}\ \bibnamefont {Martens}},\
  }\bibfield  {title} {\enquote {\bibinfo {title} {Chaos in {K}uramoto
  oscillator networks},}\ }\href {\doibase 10.1063/1.5041444} {\bibfield
  {journal} {\bibinfo  {journal} {Chaos}\ }\textbf {\bibinfo {volume} {28}},\
  \bibinfo {pages} {071102} (\bibinfo {year} {2018})}\BibitemShut {NoStop}%
\bibitem [{\citenamefont {Cheng}\ \emph {et~al.}(2017)\citenamefont {Cheng},
  \citenamefont {Guo}, \citenamefont {Dai}, \citenamefont {Li},\ and\
  \citenamefont {Yang}}]{cheng}%
  \BibitemOpen
  \bibfield  {author} {\bibinfo {author} {\bibfnamefont {H.}~\bibnamefont
  {Cheng}}, \bibinfo {author} {\bibfnamefont {S.}~\bibnamefont {Guo}}, \bibinfo
  {author} {\bibfnamefont {Q.}~\bibnamefont {Dai}}, \bibinfo {author}
  {\bibfnamefont {H.}~\bibnamefont {Li}}, \ and\ \bibinfo {author}
  {\bibfnamefont {J.}~\bibnamefont {Yang}},\ }\bibfield  {title} {\enquote
  {\bibinfo {title} {Collective chaos and period-doubling bifurcation in
  globally coupled phase oscillators},}\ }\href {\doibase
  10.1007/s11071-017-3585-z} {\bibfield  {journal} {\bibinfo  {journal}
  {Nonlinear Dyn.}\ }\textbf {\bibinfo {volume} {89}},\ \bibinfo {pages}
  {2273–2281} (\bibinfo {year} {2017})}\BibitemShut {NoStop}%
\bibitem [{\citenamefont {{Sacré}}\ and\ \citenamefont
  {{Franci}}(2016)}]{sacre16}%
  \BibitemOpen
  \bibfield  {author} {\bibinfo {author} {\bibfnamefont {P.}~\bibnamefont
  {{Sacré}}}\ and\ \bibinfo {author} {\bibfnamefont {A.}~\bibnamefont
  {{Franci}}},\ }\bibfield  {title} {\enquote {\bibinfo {title} {Singularly
  perturbed phase response curves for relaxation oscillators},}\ }in\
  \href@noop {} {\emph {\bibinfo {booktitle} {2016 IEEE 55th Conference on
  Decision and Control (CDC)}}}\ (\bibinfo {year} {2016})\ pp.\ \bibinfo
  {pages} {4680--4685}\BibitemShut {NoStop}%
\bibitem [{\citenamefont {Sherwood}\ and\ \citenamefont
  {Guckenheimer}(2010)}]{sherwood}%
  \BibitemOpen
  \bibfield  {author} {\bibinfo {author} {\bibfnamefont {W.~E.}\ \bibnamefont
  {Sherwood}}\ and\ \bibinfo {author} {\bibfnamefont {J.}~\bibnamefont
  {Guckenheimer}},\ }\bibfield  {title} {\enquote {\bibinfo {title} {Dissecting
  the phase response of a model bursting neuron},}\ }\href {\doibase
  10.1137/090773519} {\bibfield  {journal} {\bibinfo  {journal} {SIAM J. Appl.
  Dyn. Syst.}\ }\textbf {\bibinfo {volume} {9}},\ \bibinfo {pages} {659--703}
  (\bibinfo {year} {2010})}\BibitemShut {NoStop}%
\bibitem [{\citenamefont {{van Vreeswijk}}(1996)}]{Vre96}%
  \BibitemOpen
  \bibfield  {author} {\bibinfo {author} {\bibfnamefont {C.}~\bibnamefont {{van
  Vreeswijk}}},\ }\bibfield  {title} {\enquote {\bibinfo {title} {Partial
  synchronization in populations of pulse-coupled oscillators},}\ }\href
  {\doibase 10.1103/PhysRevE.54.5522} {\bibfield  {journal} {\bibinfo
  {journal} {Phys. Rev. E}\ }\textbf {\bibinfo {volume} {54}},\ \bibinfo
  {pages} {5522--5537} (\bibinfo {year} {1996})}\BibitemShut {NoStop}%
\bibitem [{\citenamefont {Politi}\ and\ \citenamefont
  {Rosenblum}(2015)}]{PR15}%
  \BibitemOpen
  \bibfield  {author} {\bibinfo {author} {\bibfnamefont {A.}~\bibnamefont
  {Politi}}\ and\ \bibinfo {author} {\bibfnamefont {M.}~\bibnamefont
  {Rosenblum}},\ }\bibfield  {title} {\enquote {\bibinfo {title} {Equivalence
  of phase-oscillator and integrate-and-fire models},}\ }\href {\doibase
  10.1103/PhysRevE.91.042916} {\bibfield  {journal} {\bibinfo  {journal} {Phys.
  Rev. E}\ }\textbf {\bibinfo {volume} {91}},\ \bibinfo {pages} {042916}
  (\bibinfo {year} {2015})}\BibitemShut {NoStop}%
\end{thebibliography}
%
%

\end{document}